# Heavily doped zinc oxide with plasma frequencies in the telecommunication wavelength range

*Alexander Koch[1]\*, Hongyan Mei[2], Jura Rensberg[1], Martin Hafermann[1], Jad Salman[2], Chenghao Wan[2,5], Raymond Wambold[2], Daniel Blaschke[3], Heidemarie Schmidt[1,3], Jürgen Salfeld[4], Sebastian Geburt[4], Mikhail A. Kats[2,5,6], and Carsten Ronning[1]*

E-mail: alexander.ak.koch@uni-jena.de

[1]Institute for Solid State Physics, Friedrich Schiller University Jena, 07743 Jena, Germany

[2]Department of Electrical and Computer Engineering, University of Wisconsin – Madison, Madison, Wisconsin 53706, USA

[3]Leibniz Institute of Photonic Technology, 07745 Jena, Germany

[4]Innovavent GmbH, 37077 Göttingen, Germany

[5]Department of Materials Science and Engineering, University of Wisconsin – Madison, Madison, Wisconsin 53706, USA

[6]Department of Physics, University of Wisconsin – Madison, Madison, Wisconsin 53706, USA

**Abstract**

We demonstrate heavy and hyper doping of ZnO by a combination of gallium (Ga) ion implantation using a focused ion beam (FIB) system and post-implantation laser annealing. Ion implantation allows for the incorporation of impurities with nearly arbitrary concentrations, and the laser-annealing process enables dopant activation close to or beyond the solid-solubility limit of Ga in ZnO. We achieved heavily doped ZnO:Ga with free-carrier concentrations of $\sim 10^{21}$ cm$^{-3}$, resulting in a plasma wavelength of 1.02 µm, which is substantially shorter than the telecommunication wavelength of 1.55 µm. Thus, our approach enables the control of the plasma frequency of ZnO from the far infrared down to 1.02 µm, thus, providing a promising plasmonic material for applications in this regime.

# 1. Introduction

Transparent conductive oxides (TCOs) have been used as transparent electrodes for solar cell applications in the past two decades,[1,2] but recently also gained attention in plasmonics and nanophotonics due to their low optical loss in the visible, metal-like behavior in the infrared, tailorable optical properties, and well-established fabrication procedures.[3-5] N-type-doped zinc oxide (ZnO) is especially attractive because its complex permittivity can be engineered over a broad wavelength range by varying the carrier concentration.[6-8] The carrier concentrations for electronics applications in the semiconductor industry are typically in the range of $10^{18} - 10^{19}$ $cm^{-3}$, corresponding to plasma frequencies in the mid infrared.[9-12] To reach plasma wavelengths in the telecom range, carrier concentrations well above $10^{20}$ $cm^{-3}$ are required, which is close to the solid-solubility limit (SSL) of n-type dopants in ZnO. Indeed, it has been demonstrated that ZnO prepared by different deposition techniques, such as magnetron sputtering or pulsed-laser deposition,[5,4] can have maximum free-carrier concentrations in the range of $(2 - 10) \times 10^{20}$ $cm^{-3}$.[13]

The SSL describes the upper limit of the concentration of impurities that are soluble in a solid at thermodynamic equilibrium for a given temperature. Impurity concentrations above the SSL typically lead to reduction of the free-carrier concentration due to multiple effects, such as secondary phase formation, dopant segregation, out diffusion, or formation of dimers, which can be electron traps.[13,14] However, doping beyond the SSL (i.e., hyper-doping) should be feasible via non-equilibrium annealing, which can suppress the processes that reduce the carrier concentration.

To date, there are only a few reports about hyper-doped materials, because non-equilibrium preparation processes are required to overcome the SSL.[15-20] Hyper-doped silicon was realized by femtosecond laser irradiation in a sulfur and nitrogen atmosphere, which resulted in very rough surfaces with microstructures.[15] Another procedure for hyper-doping is the use of ion implantation and subsequent short-time annealing process. Ion implantation is a conventional technique to dope materials with nearly any element over a broad doping-level range,[16-18] but also induces damage next to the incorporation of the dopants. Post-implantation annealing treatments are necessary to remove the irradiation damage and to activate the dopants. The use of laser or flash-lamp annealing, which can anneal on the timescale of femtoseconds to microseconds, has enabled arsenic-hyperdoped silicon and gold-hyperdoped germanium.[19,20]

The SSL of dopants in ZnO strongly depends on the temperature and is rarely reported in the literature. Yoon et al.[21] presented XRD and Raman measurements on gallium (Ga) doped ZnO samples that were annealed at 1300 °C for 5 hours. They determined a SSL of 0.5 at.% (4.2 × $10^{20}$ $cm^{-3}$) for this temperature, with higher Ga concentration resulting in the formation of secondary oxide phases.[21] Sky et al.[22] performed isochronal (constant annealing time of 30 minutes) and isothermal (annealing times between 20 minutes and 5 hours) diffusion experiments for Ga dopants in single-crystalline ZnO in the temperature range of 900 – 1050 °C and measured the concentration versus depth profiles of Ga using SIMS. Their modelling of the measured Ga diffusion profiles revealed that the dominant diffusion mechanism is consistent with zinc (Zn) vacancy ($V_{Zn}$) mediation via formation and dissociation of $Ga_{Zn}V_{Zn}$ complexes with substitutional Ga at the Zn sub-lattice ($Ga_{Zn}$).[22] The temperature-dependent SSL of Ga, aluminium (Al), and indium (In) in ZnO was determined by Sky et al. with a reaction-diffusion-type model, [22-24] which is based on Fickian diffusion plus a non-linear reaction term. More theoretical background can be found in Refs. [23,25]

**Figure 1**(a) depicts the SSL for the three n-type dopants in ZnO, Ga, Al, and In, as a function of temperature.[22-24] Ga has the highest SSL over a wide temperature range, which makes it the most promising dopant for ZnO plasmonics. This is mainly connected to the small tetrahedral covalent radius and the low formation energy of $Ga_2O_3$ compared to those of $Al_2O_3$ and $In_2O_3$. [13]

Beyond the choice of the dopant for ion implantation, the annealing time is another key factor. The annealing time must be shorter than or at least similar to the thermal response time (time of a material to react to a sudden temperature change) of ZnO in order to achieve hyper-doping, so that the dopant diffusion length is less than the average distance between dopants at a given temperature to avoiding clustering.[14] **Figure 1**(b) displays the dopant diffusion length $d_l$ versus annealing time of the three n-type dopants in ZnO at 900 °C. This diffusion length was calculated using:

$$D = D_0^{\text{eff}} \times e^{\left(-\frac{E_a}{kT}\right)}, \qquad d_l = \sqrt{D \times t_{\text{ann}}} \qquad (1)$$

with the dopant diffusivity $D$, effective diffusion constant $D_0^{\text{eff}}$, activation energy $E_a$, Boltzmann constant $k$, temperature $T$, and annealing time $t_{\text{ann}}$. We extracted the values of $D_0^{\text{eff}}$ and $E_a$ for the dopants Ga, Al, and In from references [25,23,26], where they modelled the dopant diffusion profiles at different annealing temperatures. Consequently, we used values of $D_0^{\text{eff}}$ of 0.08 cm² s⁻¹ and $E_a$ of 3 ± 0.2 eV for Ga, [25] $D_0^{\text{eff}}$ = 0.04 cm² s⁻¹ and $E_a$ of 2.6 ± 0.2 eV for Al [23] and $D_0^{\text{eff}}$ = 0.04 cm² s⁻¹ and $E_a$ of 2.2 ± 0.2 eV for In.[26] The dopant diffusivity $D$ versus annealing time in **Figure 1**(b) was then calculated using Equation 1. Long-term annealing procedures, like furnace annealing (FA) with typical annealing times on the order of 10 minutes, result in diffusion lengths of $d_l$ (Ga) ≈ 280 nm. However, laser annealing (LA) with annealing times in the nanoseconds range strongly suppresses dopant diffusion, resulting in a calculated diffusion length below 0.001 nm, which is smaller than the lattice constants of ZnO. Therefore, we implanted Ga into ZnO and used subsequent laser annealing to achieve heavy doping and plasma wavelengths close to the important telecommunication wavelength of 1.55 µm.

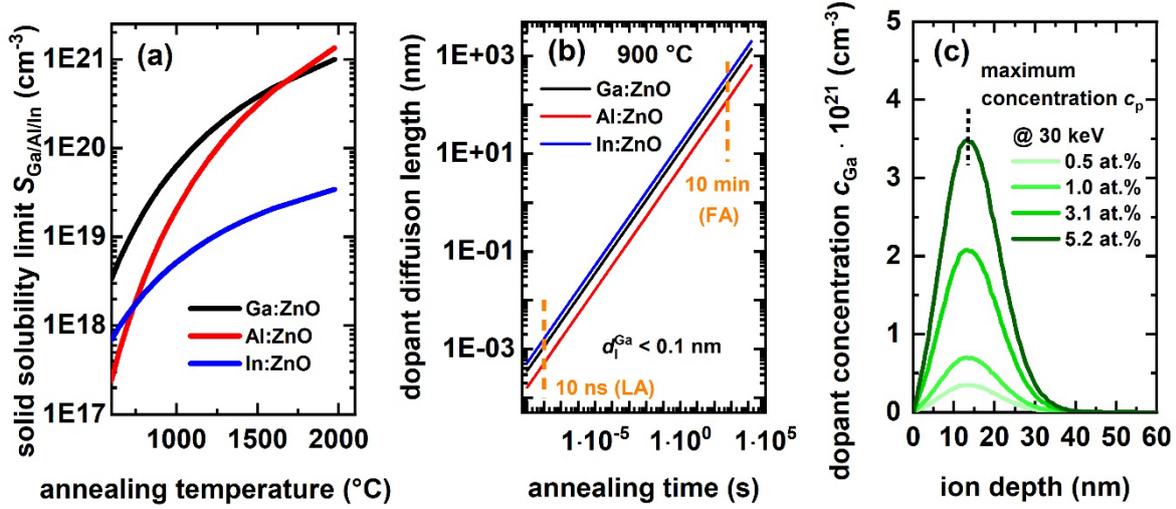

**Figure 1.** (a) Solid-solubility-limit as a function of temperature for n-type dopants (Ga, Al, In) in single-crystalline ZnO, using data taken from references.[22-24] (b) Dopant diffusion lengths for Ga, Al, and In in ZnO at 900 °C as a function of annealing time.[25,23,26] The dashed orange lines indicate typical annealing times for furnace annealing (FA) and laser annealing (LA). (c) Simulated Ga implantation profiles in ZnO using SRIM for an ion energy of 30 keV as a function of depth for different ion fluences, resulting into different gallium concentrations $c_{Ga}$.

## 2. Experimental

We irradiated single-crystalline <0001> ZnO substrates with 30 keV Ga$^+$ ions at room temperature using a commercial focused ion beam (FIB) system. The ZnO substrates are single side polished and were grown by a hydrothermal process at CrysTec® GmbH with a thickness of 0.5 mm and an area of 1 cm$^2$. Regions with sizes of 150 × 150 µm$^2$ were homogeneously irradiated. **Figure 1**(c) shows implantation profiles for different peak dopant concentrations $c_p$ as a function of depth, as simulated with the Monte-Carlo code TRIM.[27] The projected mean ion range is ≈ 14 nm and the straggle ≈ 7 nm. Each sample was implanted with ion fluences ranging from 6 × 10$^{14}$ to 6 × 10$^{15}$ cm$^{-2}$, resulting in peak values of 0.5 to 5.2 at.% that correspond to Ga concentrations of 4.28 × 10$^{20}$ to 4.28 × 10$^{21}$ cm$^{-3}$, respectively.[28]

Post-implantation annealing was performed with an INNOVAVENT VOLCANO® UV laser line beam system, which was powered by a TRUMPF TruMicro® 7370 laser with a peak laser power $P_{laser}$ of 180 W at 10 kHz, with a pulse width of 15 ns. The UV laser light ($\lambda_{laser}$ = 343 nm) was formed to a beam with a "line" cross-section, with full-width at half-maximum (FWHM) of 30 µm with Gaussian shape in the short axis and top-hat profile with a homogeneity better than 5 % (pulse peak-to-pulse peak) at a line length of 15 mm in the long axis. This line beam was scanned at a velocity of 30 mm/s over the sample surface with a 3 µm pitch, resulting in a consecutive laser spot area overlap of 90 %. The laser energy density could be set by a variable attenuator to energy densities of 100, 150, 200, 250, 300, and 350 mJ cm$^{-2}$ resulting in different annealing temperature profiles and peak annealing temperatures. The peak annealing temperatures were simulated (see Supplementary Information S1) to be about 570 °C (100 mJ cm$^{-2}$), 920 °C (150 mJ cm$^{-2}$), 1300 °C (200 mJ cm$^{-2}$), 1700 °C (250 mJ cm$^{-2}$). For laser energy densities above 250 mJ cm$^{-2}$, the peak annealing temperatures reaches the melting point of ZnO, which is 1975 °C. A first-order phase transition in a material always needs additional latent heat where the temperature does not increase while the phase change is occurring. Increasing the laser energy densities beyond 250 mJ cm$^{-2}$, the extent of melting in ZnO increases and the temperature gradient between heated regions at the surface and the bulk material increases and results in mechanical stress with destructive surface effects, like cracks or partial ablation (see Supplementary Information part 2). The onset of these destructive effects sets the maximum laser energy density for successful laser annealing of Ga-doped ZnO to about 250 mJ cm$^{-2}$. Regarding the surface quality of our Ga doped and laser annealed ZnO samples we performed atomic force microscopy (AFM) measurements after the ion implantation and laser annealing. The surface quality is represented by the root mean square (RMS) surface roughness for an area of 2 μm x 2 μm. A maximum $R_{RMS}$ value of 1.8 nm was measured for Ga doped ZnO samples, which were highly doped ($c_p$ = 5.2 at. %) and annealed at 150 – 250 mJ cm$^{-2}$. This value is only a factor of seven higher compared to the pristine ZnO, and not orders of magnitude, and thus the surfaces can be still considered as of good quality.

The comparison of $R_{RMS}$ values from our heavily doped ZnO with as grown doped ZnO thin films grown [29,30,31] by different deposition methods shows that our highly doped ZnO samples have a better surface quality for a broad range. For detailed information please look up Supplementary Information Section 4.

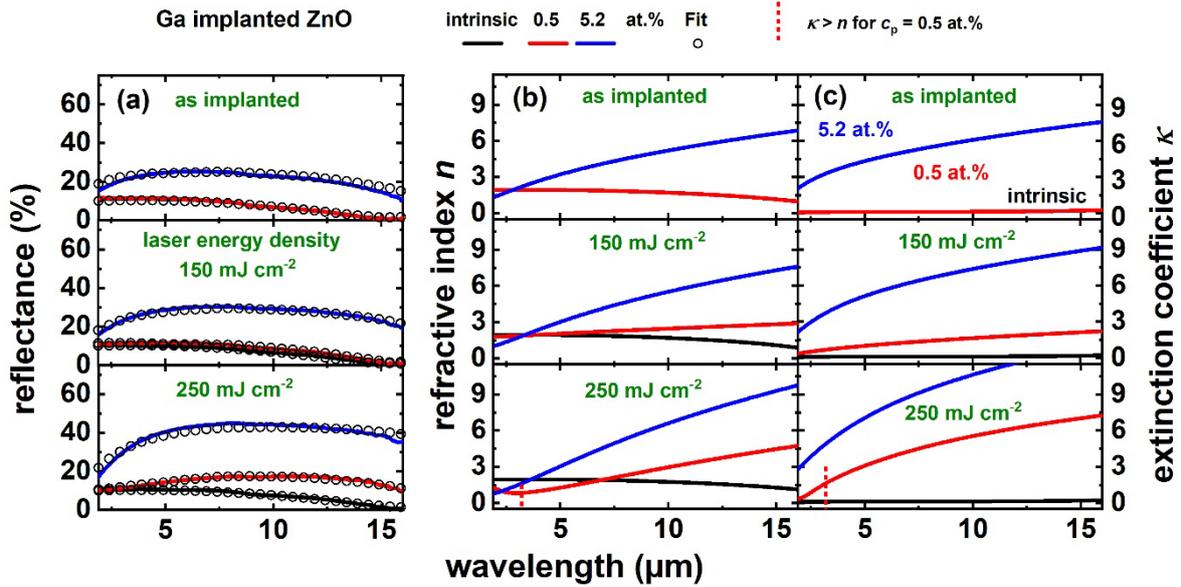

**Figure 2.** (a) Measured (solid lines) and fitted (dotted lines) reflectance spectra of intrinsic (black) and Ga-implanted ZnO (red, blue) in the unannealed state and after laser annealing with different laser energy densities (150 and 250 mJ cm$^{-2}$). (b,c) Real part ($n$) and imaginary part ($\kappa$) of the complex refractive index $\tilde{n}$ extracted from Drude-Lorentz fits to the experimental reflectance spectra from (a) for the gallium-implanted top layer. The perpendicular red dashed line in (b,c) indicates the plasma wavelengths, where $\kappa$ starting to get bigger than n for Ga implanted ZnO with $c_p$ = 0.5 at.%, which corresponds to plasmonic behavior.

## 3. Results and Discussion

**Figure 2**(a,top) shows reflectance spectra in the infrared regime obtained from two as-implanted and non-annealed samples with Ga peak concentrations $c_p$ of 0.5 and 5.2 at.%, measured using a Fourier-transform infrared (FTIR) spectrometer (Bruker Vertex 70)

connected to an infrared microscope (Hyperion 2000) in reflection mode. The spectrum of an intrinsic ZnO substrate is also shown for comparison (black), which is in top of **Figure 2**(a,b,c) similar to the reflectance of the sample with $c_p$ of 0.5 at.% (red). The reflectance increased with increasing Ga concentration in the as-implanted state (i.e., before laser annealing). This may be caused by various effects: the formation of Ga clusters at such high impurity concentrations [32,33] or implanted dopants are in substitutional position within the ZnO lattice without any annealing, which was experimentally observed for dopants in several references. [34-39]

To clarify the increasing reflectance with increasing Ga concentration in the as-implanted state, we also irradiated, with a conventional ion implanter instead of the FIB, single-crystalline ZnO with Krypton (Kr) at 30 keV and an ion fluence of $4.6 \times 10^{15}$ cm$^{-2}$, to generate a similar defect concentration distribution as for the Ga implantation with $6 \times 10^{15}$ cm$^{-2}$ (see Supplementary Information of reference [40]). The implanted, chemically inert Kr atoms do not form bonds and do not result in doping. Thus, by comparing the data of both sets of samples, we isolated the influence of implantation-induced defects on the reflectance. The reflectance of the Kr-implanted samples did not change with respect to the increasing ion fluence and is comparable to that of the pristine ZnO sample (see Supplementary Information of reference [40]). Thus, we infer that the defects and damage due to ion implantation are not responsible for the strong increasing reflectance in the infrared for the Ga-implanted samples.

The difference in reflectance for Kr and Ga implantation is likely connected to chemical and electronic doping by the implanted Ga, which is not the case for Kr. Gallium acts as a donor on Zn sub-lattice sites, and gallium clusters or charged defect complexes like $Zn_{Ga}V_{Ga}^-$ could be also formed, which all can increase the free-carrier concentration resulting in an increase in infrared reflectance without annealing. The substitutional position of implanted dopants on Zn-sites in ZnO directly after implantation at room temperature has been found experimentally for many dopants (i.e., Fe, [34] Cu, [35] Ag, [36] Er, [37,38] and Tm[39]) by Rutherford backscattering/channeling spectrometry and emission channeling spectroscopy. Note that, typically, ZnO is not amorphized upon such implantation conditions, with ion fluencies lower than $10^{16}$ cm$^{-2}$. [41,42]

The optical constants used to describe the intrinsic and unannealed Ga doped ZnO substrate were extracted from ellipsometry measurements, published in the SI of reference [40] and also presented for the as implanted Ga doped ZnO in the Supplementary Information Section 5. To quantitatively study the optical properties, i.e., the complex refractive index, of the heavily doped ZnO, we built a two-layer optical model to fit our reflectance measurements: a semi-infinite substrate of intrinsic ZnO and a top layer of highly doped ZnO:Ga, for which the optical properties can be well described using Drude-Lorentz oscillation functions: [43-47]

$$\varepsilon(\omega) = \varepsilon_D(\omega) + \varepsilon_{Ph}(\omega) = \varepsilon_\infty - \frac{\omega_p^2}{(\omega^2 + i\Gamma_D \omega)} + \Delta\varepsilon_1 \times \frac{(\omega_{Ph}^2 - i\gamma^*\omega)}{(\omega_{Ph}^2 - \omega^2 - 2i\omega\Gamma_{Ph})}. \quad (2)$$

We used Lorentzian oscillators to model the vibrational modes in the infrared, which we expect to be present in both intrinsic ZnO and in ZnO:Ga. All parameters of the Lorentzian function were taken from references [48-53] and set as constants during fitting. We used the Drude term to estimate the free-electron contribution due to substitutional Ga+ doping. We set the screened plasma frequency $\omega_p$ and damping factor $\Gamma_D$ of the Drude model to be the fitting parameters, which were swept to find a reflectance that matched with our FTIR measurements. The fitting parameters $\omega_p$ and $\Gamma_D$ on the reflectance are crucial for further data processing to calculate the free carrier concentration $N$, plasma wavelength $\lambda_p$ and dopant activation efficiency (DAE). The parameters $\omega_p$ and $\omega_N$ are known as frequencies of the plasma resonance and defined as:

$$\omega_p^2 = \frac{Ne^2}{\varepsilon_0 m^*}, \quad \omega_N^2 = \frac{Ne^2}{\varepsilon_\infty \varepsilon_0 m^*} \qquad (3)$$

Note that the relation between $\omega_p$ and $\omega_N$ is $\omega_N^2 = \omega_p^2/\varepsilon_\infty$, where $\omega_p$ is named in some literature as screened plasma frequency.

The plasma wavelength $\lambda_p$ can be calculated with the relation between energy of electromagnetic radiation and their wavelength:

$$E = \frac{h \cdot c}{\lambda}, \quad \lambda_p \text{ (nm)} = \frac{1239.83}{E \text{ (eV)}} \qquad (4)$$

where $c$ is the speed of light and $h$ the Planck's constant. Using the Equations 3 and 4, we determined the free carrier concentration $N$ of our samples. The dopant activation efficiency (DAE) describes the ratio between activated free carrier concentration and the average incorporated dopant concentration for a depth of 30 nm, obtained from the SRIM calculated Gaussian ion implantation profile, see **Figure 1**(c). The average dopant concentration is calculated by an integral over the gallium concentration profile over sample depth and divided by the 30 nm thickness of the Ga doped ZnO layer. For detailed information see Supplementary Information S3 and S5.

For the layer of Ga-implanted ZnO, we assumed a homogeneous film with a thickness of 30 ± 10 nm, which covers 97 % of the implanted Ga profile, according to TRIM simulations shown in **Figure 1**(c). This assumption for the thickness is also well justified by the anticipated lack of diffusion during the annealing process As shown in **Figure 2**(a), the good agreement between our models (dotted lines) and reflectance measurements (solid curves) enabled us to exact the real and imaginary parts of the complex refractive index of the resulting ZnO:Ga (**Figure 2**(b,c)).

A characteristic for Drude-metal-like behavior for a material is when the extinction coefficient $Im\{\tilde{n}\} = \kappa$ becomes higher than the real part of the refractive index $Re\{\tilde{n}\} = n$, i.e., $\kappa > n$,[54] like in **Figure 2**(b,c) this is highlighted with a perpendicular red dashed line, which indicates that Ga-implanted ZnO has a gallium concentration peak $c_P$ of 0.5 at.%. At this point, we get a plasmonic material, which could be useful for plasmonic applications in the near infrared.

We performed laser annealing at a laser energy density of 100 mJ cm$^{-2}$, but did not observe any significant changes in the reflectance signal compared to the as implanted state, see Supplementary Information S2. However, reflectance starts to slightly increase at a laser energy density of 150 mJ cm$^{-2}$ due to dopant activation (**Figure 2**(a)). This dopant activation appears to be more effective for low gallium ion fluences. However, stable defect clusters are formed for higher ion fluences, which require higher annealing temperatures for healing.[29,30] An increase of the laser power results into a further increase of the reflectance. For the investigated laser energy densities of 100 - 350 mJ cm$^{-2}$, the maximum reflectance signal is achieved using laser energy densities of 250 mJ cm$^{-2}$, which is shown in at the bottom of **Figure 2**(a). This increase can be attributed to the electrical and optical activation of Ga dopants in the ZnO lattice, thus the free-carrier concentration increases. A further increase of the laser energy density to 300 mJ cm$^{-2}$ and above resulted in degradation of the sample, which we observed with a strong increase of the reflectance of our intrinsic ZnO reference sample. The reflectance increase is caused by the sublimation of the surface layer, where the oxygen diffuses out and a Zn-enriched surface layer forms. Via SEM, we investigated cracks through the ZnO and partly ablated portions of the material, see supplementary information part S2.

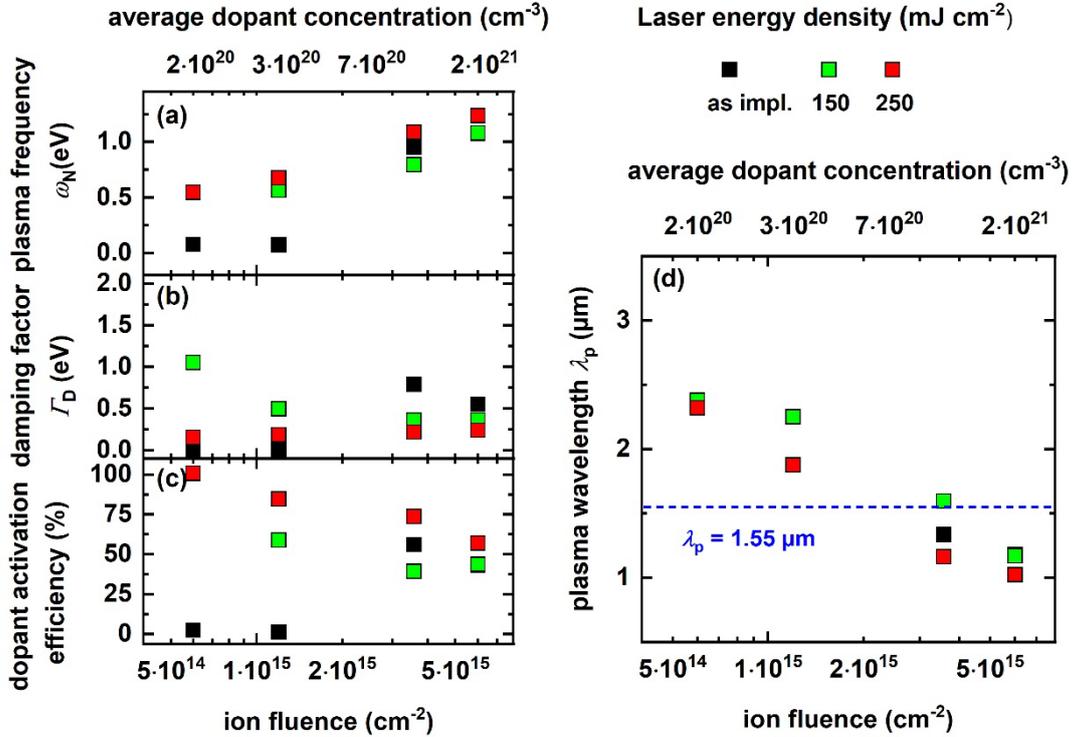

**Figure 3.** (a) Plasma frequency $\omega_N$ and (b) damping factor $\Gamma_D$ extracted from the fits shown in **Figure 2** as a function of the ion fluences for different laser energy densities. (c) Dopant activation efficiency, which is the ratio of activated free carriers to the averaged implanted Ga concentration across the thickness, versus ion fluences. (d) Plasma wavelength $\lambda_p$ versus ion fluence. The telecommunication wavelength is indicated with a blue dashed line.

**Figure 3**(a,b) shows the plasma frequency $\omega_N$ and damping factor $\Gamma_D$ as a function of the ion fluence, where $\omega_N$ and $\Gamma_D$ were extracted from the fits to reflectance spectra in **Figure 2.** Details and the complete data sets are present in the Supplementary Information S2. Increasing both the ion fluence and the laser energy density usually leads to an increase of $\omega_N$, and the highest plasma frequencies of 1-1.3 eV were obtained for a laser energy density of 250 mJ cm$^{-2}$ and an ion fluence range of $3.6 \times 10^{15} - 6 \times 10^{15}$ cm$^{-2}$ (equivalently, peak dopant concentration of 3.1 – 5.2 at.%). Degradation of the ZnO samples occurs beyond this value, as mentioned above. The damping factor is inversely proportional to the carrier mobility and **Figure 3**(b) indicates that higher dopant concentrations result into a lower mobility, which is consistent with increased electron-defect scattering.

**Figure 3**(c) depicts the dopant activation efficiency, which is the ratio of the achieved free-carrier concentration for a given averaged implanted Ga concentration across the thickness of 30 nm. Laser annealing with 150 mJ cm$^{-2}$ results in activation of 95 % to 40 % of the incorporated Ga atoms with increasing ion fluence, whereas values of 100-50 % can be achieved by laser annealing using 250 mJ cm$^{-2}$. However, the activation efficiency also decreases with increasing dopant concentration, which is most probably limited by the temperature-dependent SSL of Ga in ZnO.

Furthermore, we extracted the activated dopant concentration from the fits and plotted in **Figure 3**(d) the respective plasma wavelength, defined as $\lambda_p = \omega_p/2\pi$, versus the used ion fluence range for all investigated ZnO:Ga layers. We obtained the highest carrier concentration of $(1.23 \pm 0.41) \times 10^{21}$ cm$^{-3}$ for an ion fluence of $6 \times 10^{15}$ cm$^{-2}$ and for laser annealing with a laser energy density of 250 mJ cm$^{-2}$. This value corresponds to a plasma wavelength of $1.02 \pm 0.17$ μm, which is below the telecommunication wavelength of 1.55 μm (blue dashed line in **Figure 3**(d)). This value translates also to an activation of $\sim 10^{21}$ cm$^{-3}$ Ga atoms in the ZnO lattice, which is close to the SSL of Ga at a temperature of 1975 °C (compare **Figure 1**(a)).

**Figure 4** highlights the SSL of Ga compared to the activated dopant concentrations of our ZnO:Ga samples, irradiated with an ion fluence of $6 \times 10^{15}$ cm$^{-2}$ ($c_P$ = 5.2 at.%), for the investigated temperature range. Here, the laser energy has been converted to a temperature using the material parameters of ZnO, the Beer-Lambert law, and the heat equation (see supplementary information S1). For the laser energy density range of 150 mJ cm$^{-2}$ – 250 mJ cm$^{-2}$ we achieved hyper-doped ZnO. Increasing the laser energy density towards the melting temperature of ZnO led to values close to the SSL of Ga in ZnO. [22]

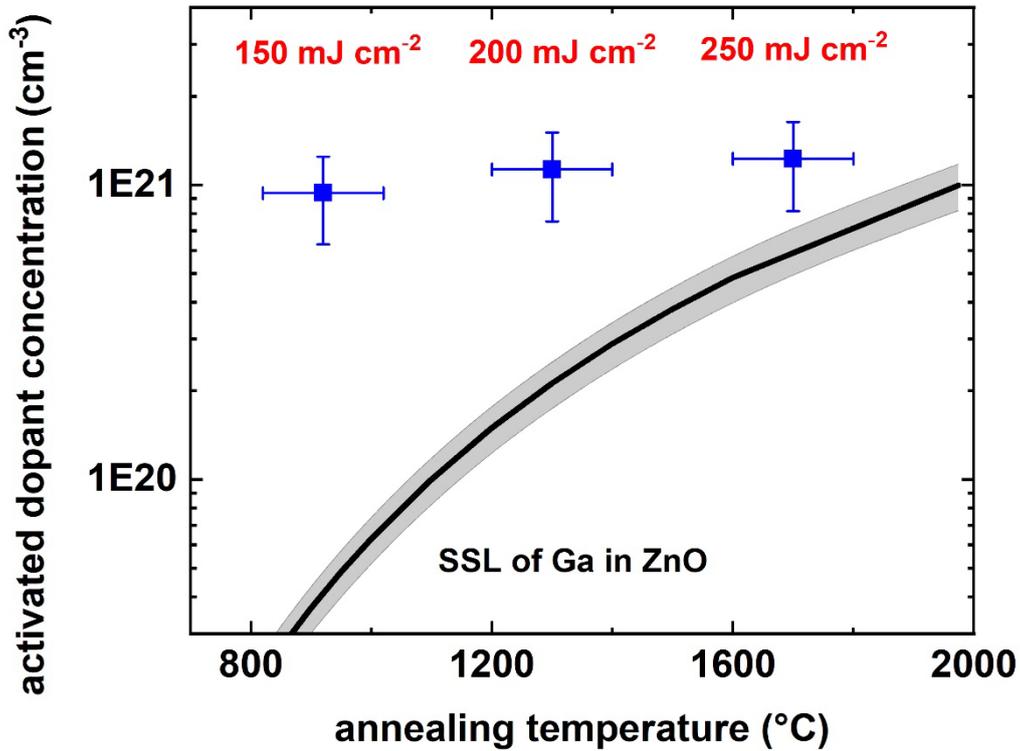

**Figure 4.** Comparison between activated free-carrier concentrations, from the Ga doped samples irradiated with an ion fluence of $6 \times 10^{15}$ cm$^{-2}$ ($c_P$ = 5.2 at.%) and laser annealed in the laser energy density range 150 mJ cm$^{-2}$ – 250 mJ (corresponds to an annealing temperature range 920 °C and 1700 °C) and the temperature-dependent solid solubility limit of Ga in ZnO taken from the work of Sky et al [22], where the grey area around the black line displays an error of ± 18 % for the solid solubility limit.

## 4. Conclusion

In conclusion, we gradually tuned the optical properties of ZnO by ion implantation of gallium and subsequent nanosecond laser annealing. Using this approach, we achieved even plasma wavelengths in the near-infrared spectral region, including for wavelengths shorter than 1.55 µm. The highest achieved free-carrier concentration in ZnO:Ga was $N_e \approx 1.23 \pm 0.41 \times 10^{21}$ cm$^{-3}$, corresponding to $\lambda_p = 1.02 \pm 0.17$ µm, which is 1.48 at. % substituted Gallium in the ZnO host lattice with a doping level of $2.16 \times 10^{21}$ cm$^{-3}$ (2.6 at. %) and a dopant activation efficiency of 57 %. In general, with this doping approach we are reaching dopant activation efficiencies in the range from 30 % up to 100 %. This is enabled by the combination of high concentration ion implantation and the high peak annealing temperatures during the short laser annealing in the nanosecond regime, where dopant diffusion is suppressed, which is advantageous for maintaining spatially localized distributions of dopants, resulting even in hyper-doping of ZnO under certain conditions. Heavily or hyper-doped ZnO:Ga with plasma frequencies in the telecommunication range range have a broad applicability. With a focused ion beam we are able to selective and gradually dope ZnO. Thus, it opens a wide field for lateral arrangements of sub-wavelength sized and spaced optical doped and un-doped regions with a high optical contrast, which are a promising approach for photonic or plasmonic applications in the NIR spectral range, for example with an area selective ion implantation process

## Acknowledgements

We thank the German Research Foundation (DFG – Deutsche Forschungsgemeinschaft) for funding this work via grant Ro1198/21-1 as well as Office of Naval Research (Grant # N00014-20-1-2297) and also the financial support by the Thüringer Aufbaubank (EFRE-OP 2014-2020). The UW-Madison authors acknowledge use of facilities and instrumentation at the UW-Madison Wisconsin Centers for Nanoscale Technology (wcnt.wisc.edu) partially supported by the NSF through the University of Wisconsin Materials Research Science and Engineering Center (DMR-1720415).

# Heavily doped zinc oxide with plasma frequencies in the telecommunication wavelength range

*Alexander Koch[1\*], Hongyan Mei[2], Jura Rensberg[1], Martin Hafermann[1], Jad Salman[2], Chenghao Wan[2,5], Raymond Wambold[2], Daniel Blaschke[3], Heidemarie Schmidt[1,3], Jürgen Salfeld[4], Sebastian Geburt[4], Mikhail A. Kats[2,5,6], and Carsten Ronning[1]*

E-mail: alexander.ak.koch@uni-jena.de

[1]Institute for Solid State Physics, Friedrich Schiller University Jena, 07743 Jena, Germany

[2]Department of Electrical and Computer Engineering, University of Wisconsin – Madison, Madison, Wisconsin 53706, USA

[3]Leibniz Institute of Photonic Technology, 07745 Jena, Germany

[4]Innovavent GmbH, 37077 Göttingen, Germany

[5]Department of Materials Science and Engineering, University of Wisconsin – Madison, Madison, Wisconsin 53706, USA

[6]Department of Physics, University of Wisconsin – Madison, Madison, Wisconsin 53706, USA

## 1. Laser Annealing Process and Simulation of Temperature Profile

To obtain information about the temperature distribution during short term annealing, like laser annealing, the simulation of time-dependent temperature profiles has been performed in some reports. [1-3] The fraction of light absorbed by the single-crystalline ZnO and the corresponding absorption depth profile is determined by the optical properties of the ZnO. A numerical simulation with MATLAB, based on the following Equation S1 and S2, was implemented and produces the temperature profile versus sample depth. The temperature distribution is obtained numerically by solving the heat equation [1-3]:

$$\rho(T)c_\mathrm{p}(T)\frac{\partial T(x,t)}{\partial t} - \frac{\partial}{\partial x}\left(\kappa(T)\frac{\partial T(x,t)}{\partial x}\right) = Q(x,t) \quad (S1)$$

here $\rho(T)$ is the mass density, $c_\mathrm{p}(T)$ is the specific heat capacity, and $\kappa(T)$ is the thermal conductivity, where the parameters are: $\rho(T)$ = 5.7 g cm$^{-3}$, $c_\mathrm{p}(T)$ = 0.494 J g$^{-1}$ K$^{-1}$ and $\kappa(T)$ = 2.5 W m$^{-1}$ K$^{-1}$.[4] The left side represents the temporal and spatial variation of the temperature and of the temperature-dependent materials properties. [1-3]

The right side $Q(x,t)$ describes how radiation is absorbed by the material according to the Beer – Lambert law.[1-3] For a semi-infinite material with homogeneous optical properties, $Q(x,t)$ can be written:

$$Q(x,t) = I_\text{L} \cdot (1-R) \cdot \alpha(T) \cdot e^{(-\alpha(T)x)} \qquad \text{(S2)}$$

here $I_\text{L}$ is the energy density of the laser pulse, here in the used range between 100 – 300 mJ cm$^{-2}$. The reflectance $R$ and transmittance $T$ of pristine ZnO was measured with an UV-VIS spectrometer at room temperature in a spectral range of 300 – 2500 nm. At the laser wavelength of 343 nm $R = 0.148$ and $\alpha(T) = 4.3 \times 10^4$ cm$^{-1}$.

The resulting temperature profile versus sample depth for each laser density is shown in **Figure S1**. The peak annealing temperature increases with increasing laser density. With laser densities higher than 250 mJ cm$^{-2}$, the peak annealing temperature in the first 40 ns is close to the melting temperature of ZnO (1975 °C). Thus, for the laser densities of 300 mJ cm$^{-2}$ and 350 mJ cm$^{-2}$, the single crystalline ZnO is at the transformation between solid and liquid phases, like a mixed-phase regime. We expect at this phase transition, a small increase of the temperature profile with laser energy density, which is connected to the heat of transformation. The heat of transformation is energy absorbed by ZnO during the laser annealing for a phase change in ZnO. Furthermore, the very fast heat input from room temperature to peak annealing temperature and back to 1000 K within ZnO happens in the first 100 ns. The temperature stays at the peak annealing temperature for only a few ns.

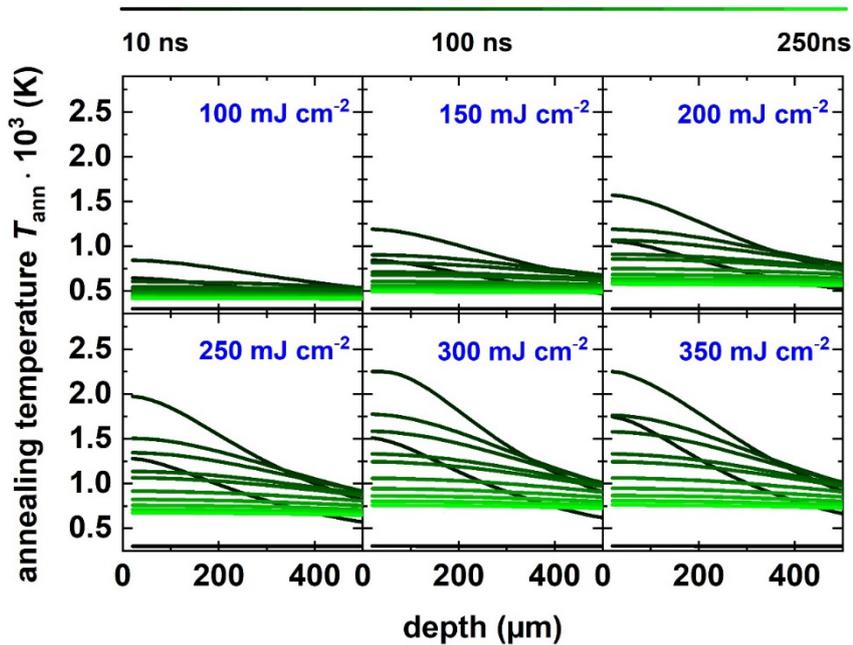

**Figure S1.** Simulated annealing temperature profiles during the laser annealing process at different laser densities as a function of sample depth and for a simulation time of 250 ns.

## 2. FTIR measurement in reflectance mode and Drude-Lorentz fit results of ZnO:Ga

All reflectance measurements of Ga doped single crystalline ZnO are given in **Figure S2**. The reflectance was measured at normal incidence with FTIR, which simplifies the Fresnel equations for one interface. In case of Gallium doped ZnO, we expect a complex refractive index, where the extinction coefficient is $\kappa > 0$. The reflectance for laser annealed intrinsic ZnO strongly increases with laser energy densities $\geq 300$ mJ cm$^{-2}$ compared to the reflectance at lower laser energy densities between 100 and 250 mJ cm$^{-2}$, which are yet comparable to as-implanted ZnO. This effect is most likely caused by local melting with annealing temperatures close and above to the ZnO melting point. Mainly a thin layer at the surface of the ZnO substrate faces the high peak annealing temperatures, whereas the backside and most of the bulk remain at room temperature. [1] An explanation could be an out diffusion of vaporized oxygen caused by laser annealing temperature leading to an enriched metallic Zinc surface layer, which indeed would increase the reflectance. Furthermore, the reflectance still increases for low gallium fluences at 300 mJ cm$^{-2}$, where we most probably have a process combination of an enriched Zn surface layer and still electrical dopant activation. At high doping levels with $c_p$ of 3.1 and 5.2 at.% the reflectance slightly decreases, because above the SSL the probability of oxide phase formation and acceptor-like defect formation is very high. They are electrical active like electron traps and reducing the free carrier concentration. A laser density of 350 mJ cm$^{-2}$ leads to strong destructive effects like cracks and material ablation for all samples, where the destructive impact increases with increasing gallium concentration, which is not shown in **Figure S2**.

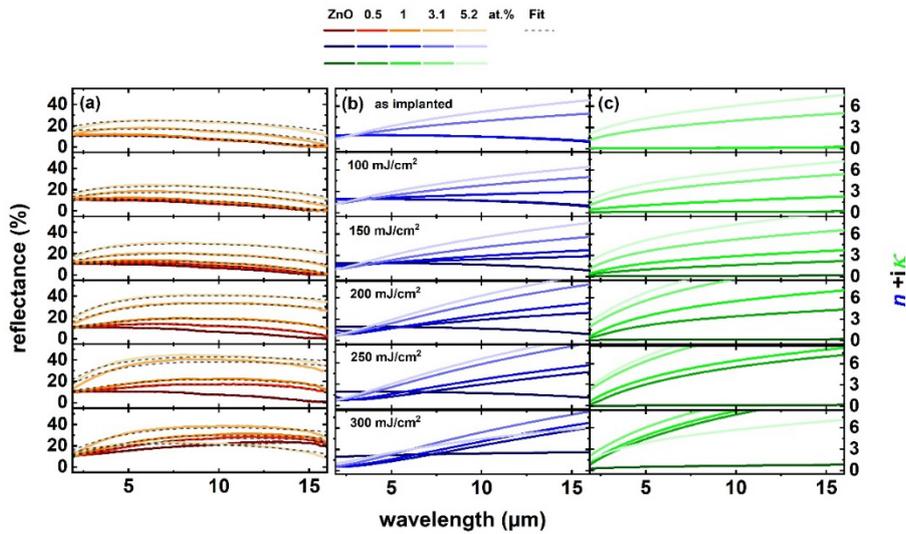

**Figure S2.** (a) Reflectance of Ga doped ZnO substrates at laser densities between 100 – 300 mJ cm$^{-2}$ and all Gallium ion fluences plus the real and imaginary part of the extracted complex refractive indices in column (b) and (c).

**Figure S3** shows the results of the Drude-Lorentz fit routine extracted from the reflectance spectra in **Figure S2** for the laser energy density range 100 – 250 mJ cm$^{-2}$ and the as implanted state. The error bars in **Figure S3** are estimated with the thickness variation within the fit routine of the thin Ga doped ZnO layer with first 20 nm, then 30 nm and finally with 40 nm during the simulation.

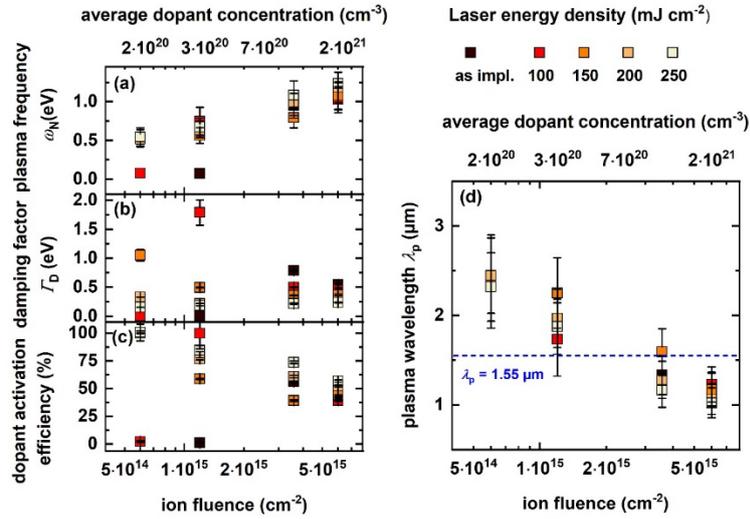

**Figure S3.** Complete data set of the extracted fitting parameters plasma frequency and damping factor from Drude-Lorentz fit routine of the unannealed and laser annealed (100 – 250 mJ cm$^{-2}$) ZnO:Ga samples. The error bars are estimated with a thickness variation of the high Ga doped ZnO layer from 20 nm, 30 nm and 40 nm.

**Figure S4** presents SEM pictures of ion irradiated and laser annealed ZnO:Ga substrates. The transition from a laser density of 250 mJ cm$^{-2}$ to 300 mJcm$^{-2}$ marks a huge increase in destructive annealing effects, which can be clearly identified in **Figure S4**(b). All samples that were laser annealed with laser energy densities below 250 mJ cm$^{-2}$ do not show any obvious degradation.

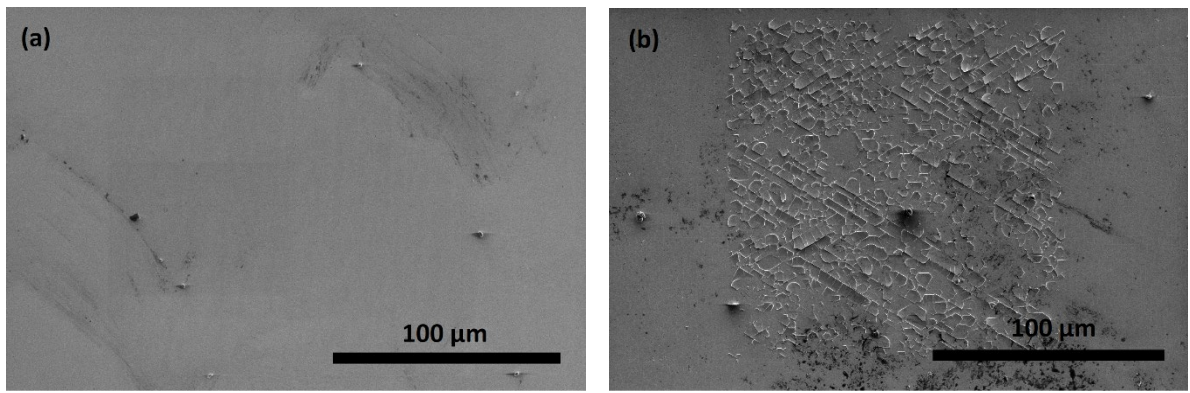

**Figure S4.** SEM-pictures from irradiated ($c_p$ = 5.2 at.%) and laser annealed Ga:ZnO samples at 250 mJ cm$^{-2}$ in (a) and 300 mJ cm$^{-2}$ in (b). The surface degradation after laser annealing above 250 mJ cm$^{-2}$ is obvious.

**Figure S4**(b) shows compared to **Figure S4**(a) partially laser melting, cracks, and a rougher surface within the implantation region. Because of these destructive effects, we exclude all Ga doped ZnO for data evaluation, which were annealed above 250 mJ cm$^{-2}$.

3. **Drude-Lorentz fit to reflectance spectra of Ga (@ 30 keV) implanted single crystalline ZnO**

Simulated reflectance spectra, calculated with Equation S13 were implemented in a MATLAB code and used for fitting the measured reflectance of the Ga doped ZnO. In the following the theoretical background for the calculation of the reflectance spectra is presented, which is based on several publications.[13-17] The reflectance and transmittance coefficients for s- and p-polarised light are $r_s$, $r_p$ and $t_s$, $t_p$ respectively, of a single interface between two semi-infinite media can be calculated by using the Fresnel Equation.[18,19] Note that all the values are complex numbers. In case of a thin film with a finite thickness, there are two interfaces and multiple reflections from the 1st and 2nd interface may occur. For such cases the 2 x 2 matrix method is used to calculate the total reflectance or transmittance. The resultant matrix $M_r$ for a thin film on a semi-infinite substrate has to be known, to calculate the reflectance spectra:

$$M_r = M_{mf} M_f M_{fs} \qquad (S3)$$

$M_{mf}$ and $M_{fs}$ are the interface matrices of the medium-film and film-substrate interfaces respectively and defined as:

$$M_{j,j+1} = \frac{1}{t_{j,j+1}} \begin{pmatrix} 1 & r_{j,j+1} \\ r_{j,j+1} & 1 \end{pmatrix} \qquad (S4)$$

with $j$ the medium number. $M_f$ is the propagation matrix of the thin film:

$$M_f = \begin{pmatrix} \exp(ib_f) & 0 \\ 0 & \exp(-ib_f) \end{pmatrix} \qquad (S5)$$

Here $b_f$ is the phase thickness of the thin film $f$ defined as:

$$b_f = \frac{2\pi d_f}{\lambda} n_f \cos(\varphi_f) \qquad (S6)$$

where $\lambda$ is the wavelength and $d_f$, $n_f$ and $\varphi_f$ are the thickness, the complex index of refraction of the film $f$ and $\varphi_f$ is the incidence angle of light. Knowing the resultant matrix $M_r$ the reflection coefficients can obtain from:

$$r_{p,s} = \frac{M_{r2,1}}{M_{r1,1}} \qquad (S7)$$

Finally, the total reflectance was calculated by

$$R = \frac{r_p \cdot r_p^* + r_s^* \cdot r_s}{2} \qquad (S8)$$

Here $r_p^*$ and $r_s^*$ denote the complex conjugates of the $r_s$ and $r_p$ complex numbers. The complex optical constants ($\tilde{n} = n + i\kappa$) of all media must be known in order to simulate the measured reflectance spectra. If the dielectric function $\tilde{\varepsilon} = \varepsilon_1 + i\varepsilon_2$ is known, we can calculate the optical constants with the relationship $\sqrt{\tilde{\varepsilon}(\omega)} = n + i\kappa = \tilde{n}$ for non-magnetic and isotropic materials. The real part of the complex refractive index is the index of refraction $Re\ \{\tilde{n}\} = n$ and the extinction coefficient is the imaginary part $Im\ \{\tilde{n}\} = k$. The optical constants of the pristine ZnO substrate were extracted by spectral ellipsometry measurement, data taken from [20].

To describe the optical properties of the Ga doped ZnO thin film we used a Drude-Lorentz model. A Drude model is used to describe the optical properties of the sample when the plasma resonance wavelength is far away from lattice vibrations. The classical Drude model can be written in the form:

$$\varepsilon_D(\omega) = \varepsilon_\infty - \frac{\omega_p^2}{(\omega^2 + i\Gamma_D\omega)} = \varepsilon_\infty \left(1 - \frac{\omega_N^2}{(\omega^2 + i\Gamma_D\omega)}\right) \quad (S9)$$

where $\varepsilon_\infty$ high-frequency dielectric constant (we used $\varepsilon_\infty = 3.63$ F m$^{-1}$ for ZnO as a fixed parameter for all samples) and $\Gamma_D$ is the damping factor. Both parameter $\omega_p$ and $\omega_N$ are known as frequency of the plasma resonance and defined as:

$$\omega_p^2 = \frac{Ne^2}{\varepsilon_0 m^*}, \quad \omega_N^2 = \frac{Ne^2}{\varepsilon_\infty \varepsilon_0 m^*} \quad (S10)$$

Note the relation between $\omega_p$ and $\omega_N$ is $\omega_N^2 = \omega_p^2/\varepsilon_\infty$, where $\omega_p$ is named in some literature as screened plasma frequency.

The fitting parameters $\omega_p$ and $\Gamma_D$ from the Drude-Lorentz fits on the reflectance are crucial for further data processing to calculate free carrier concentration $N$, plasma wavelength $\lambda_p$ and dopant activation efficiency (DAE).

**Figure 3**(a,b) in the main text plots the plasma frequency $\omega_N$, calculated by equation above, and Drude damping term $\Gamma_D$. The plasma wavelength $\lambda_p$ in **Figure 3**(d) can be calculated with the relation between energy of electromagnetic radiation and their wavelength:

$$E = \frac{h \cdot c}{\lambda}, \quad \lambda_p \text{ (nm)} = \frac{1239.83}{E \text{ (eV)}} \quad (S11)$$

where $c$ is the speed of light and $h$ the Planck's constant. Using the Equations S10 and S11, we determined the free carrier concentration $N$ of our samples. In **Figure 3**(c) the dopant activation efficiency (DAE) is plotted versus the used ion fluence. The DAE describes the ratio between activated free carrier concentration and the average incorporated dopant concentration for a depth of 30 nm, obtained from the SRIM calculated Gaussian ion implantation profile, see **Figure 1**(c) in the main text. The average dopant concentration is calculated by an integral over the gallium concentration profile over sample depth and divided by the thickness of the Ga doped ZnO layer.

In case the plasma frequency is near or lower than lattice vibrations, which is the case for pristine ZnO, the phonons have to be taken into account in the model for the optical constants. The modelling of phonon is usually made by a Lorentz oscillator:

$$\varepsilon_{Ph}(\omega) = \varepsilon_\infty + \Delta\varepsilon_1 \times \frac{(\omega_{Ph}^2 - i\gamma^*\omega)}{(\omega_{Ph}^2 - \omega^2 - 2i\omega\Gamma_{Ph})} \quad (S12)$$

where $\omega_{Ph}$ $is$ the frequency of the expected phonon, the $\Gamma_{Ph}$ is broadening or damping of the oscillator and $\Delta\varepsilon_1$ is the weighting factor of the Lorentz oscillator.[14] The positions and broadening of the phonons were studied by many different techniques like FTIR, [21,22] spectroscopic ellipsometry [23,24] and Raman spectroscopy.[25,26]

Finally, the combined dielectric function for the plasma resonance and lattice vibrations can be presented as Drude-Lorentz model:

$$\varepsilon(\omega) = \varepsilon_D(\omega) + \varepsilon_{Ph}(\omega) = \varepsilon_\infty - \frac{\omega_p^2}{(\omega^2 + i\Gamma_D\omega)} + \Delta\varepsilon_1 \times \frac{(\omega_{Ph}^2 - i\gamma^*\omega)}{(\omega_{Ph}^2 - \omega^2 - 2i\omega\Gamma_{Ph})} \quad (S13)$$

which well describes the frequency dependent complex permittivity $\varepsilon(\omega)$ of the Ga implanted ZnO.

The reflectance of an undoped ZnO single crystal was measured and fitted with the Drude-Lorentz oscillator in Equation S13, regarding the calculated reflectance spectra with the *n&κ* data from our SE measurements. Position and intensity of the Lorentz oscillator match with averaged values of the phonons in literature, [14-19] therefore we fixed the phonon position to $\omega_{Ph}$ = 0.0497 eV in all our fits. The Lorentz fit parameters, see **Table S1** of the intrinsic ZnO, were kept constant in all our fits of the as implanted and laser annealed samples, thus only the Drude term was adjusted to describe the electromagnetic response of the free carriers, with Equation S12.

**Table S1.** Extracted fit parameter of the Drude-Lorentz oscillator for intrinsic single crystalline ZnO.

| | |
|---|---|
| $\varepsilon_\infty$ | 3.63 F m$^{-1}$ |
| $\omega_p$ | 0.14 eV |
| $\Gamma_D$ | 3.1 × 10$^{-3}$ eV |
| $\Delta\varepsilon_1$ | 2.41 |
| $\omega_{Ph}$ | 0.05 eV |
| $\Gamma_{Ph}$ | 0.14 eV |
| $\gamma^*$ | 0.09 eV |

## 4. AFM measurements on pristine, Ga doped and annealed ZnO samples

We performed additional atomic force microscopy (AFM) measurements on the irradiated areas to check the surface quality. **Figure S5** displays the AFM measurements on pristine single-crystalline ZnO and the Ga irradiated (0.5 at. % and 5.2 at. %) and laser annealed areas (laser energy density range between 100 – 250 mJ cm$^{-2}$). The surface quality is represented by the root mean square (RMS) surface roughness for an area of 2 μm x 2 μm.

AFM measurements were performed on a VistaScope system from Molecular Vista Inc. using gold coated AFM tips from Nanosensors (SD-NCHAu25-W) with a resonance frequency of about 250 kHz. The images were taken in tapping mode at the 2nd mechanical resonance of the cantilever at about 1.5 MHz with a tapping amplitude of 1 nm, and a setpoint of 85%. After levelling the images with a second order line by line mode in scan direction, the root mean square (RMS) roughness was determined using SurfaceWorks version 3.0 release 30. The RMS roughness $R_{RMS}$ describes the quadratic average of profile height deviations from the mean line.

For polished wafers the RMS roughness is typically in the range between 0.1 – 0.6 nm. Indeed, the AFM measurement of our pristine ZnO substrate gives a roughness of $R_{RMS}$ = 0.25 nm. With Ga irradiation and an increase in ion fluence a slight decrease in the surface roughness is observed, see **Figure S8**. In general, sputtering is dependent on the ion mass, ion fluence and ion energy. Thus, Ga ion irradiation with our used parameters increases slightly the surface quality of ZnO. Furthermore, the laser annealing process with increasing energy density does also show no changes in surface roughness over the whole energy density range.

However, **Figure S5** shows that the combination of both non-equilibrium processes results in a slight increase in surface roughness, with increasing laser energy density and implanted gallium concentration. A maximum $R_{RMS}$ value of 1.8 nm was measured for Ga doped ZnO samples, which were highly doped ($c_p$ = 5.2 at. %) and annealed at 150 – 250 mJ cm$^{-2}$. This value is only a factor of seven higher compared to the pristine ZnO, and not orders of magnitude, and thus the surfaces can be still considered as of good quality.

The comparison of $R_{RMS}$ values from our heavily doped ZnO with as grown doped ZnO thin films grown [27,28,29] by different deposition methods shows that our highly doped ZnO samples have a better surface quality for a broad range.

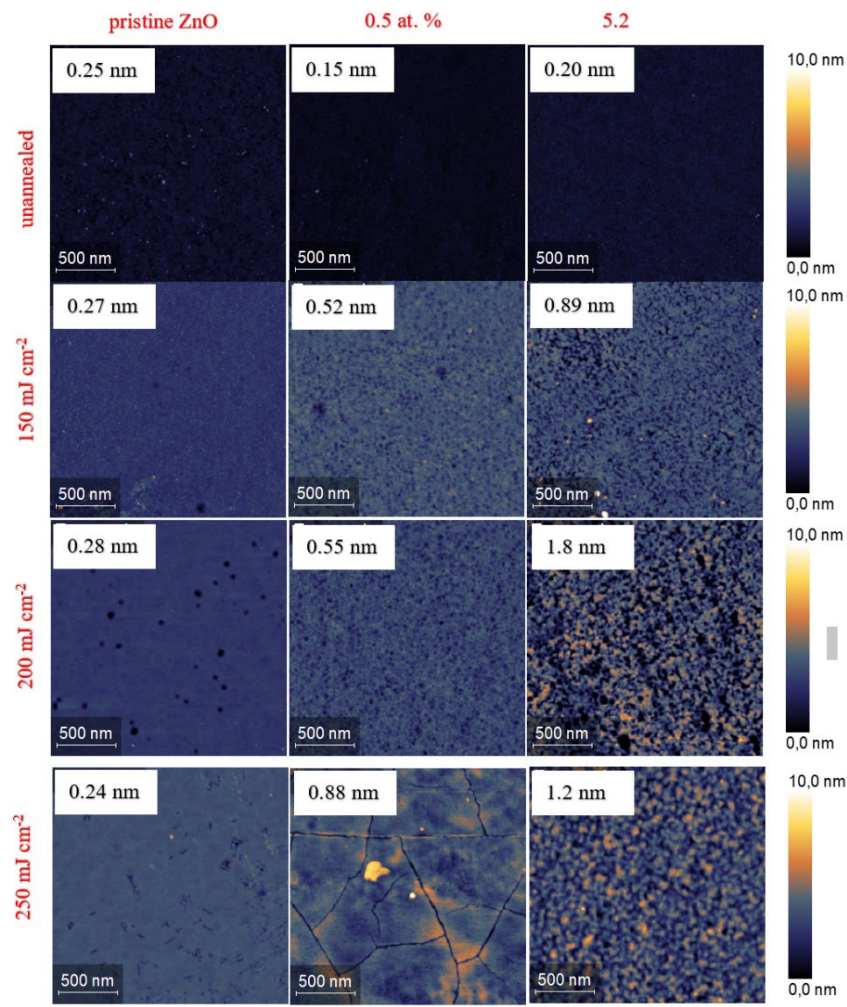

**Figure S5.** AFM measurements in tapping mode of pristine ZnO, Ga doped and laser annealed ZnO samples for an area of 2 µm x 2 µm. From top to bottom the laser energy density increase from unannealed state until 250 mJ cm$^{-2}$. From left to right the gallium doping level increases from intrinsic ZnO, 0.5 at. % and 5.2 at. %. It's obvious from the AFM surface scan that ion implantation or laser annealing don't change the surface quality of our fabricated samples. With increasing the ion fluence the surface roughness decreases slightly. An increase in laser energy density don't change anything at pristine ZnO. The combination of both non-equilibrium processes gives a rougher surface up to 1.8 nm for $R_{RMS}$.

## 5. Comparison between extracted free carrier concentration and mobility from Drude Lorentz fit and Spectral Ellipsometry measurements for unannealed Ga irradiated ZnO

The Ga ion implantation was done by a focused ion beam with an irradiated area of 200 μm x 200 μm, where in total 4 different regions are implanted with different ion fluences into one single-crystalline substrate to ensure and guarantee the exact same annealing conditions at one laser energy density. In total 6 single-crystalline ZnO substrates were irradiated, each one with 4 implanted regions and then annealed at different laser energy densities at once.

Hall effect measurements on the small implanted regions are challenging regarding preparation of ohmic contacts and to realize the electrical isolation of the irradiated area from the remaining ZnO host substrate as well as from neighboring regions. This would require costly a new design of the experiment in combination with costly electron beam lithography to define contact pads and lines. Unfortunately, spectroscopic ellipsometry measurements can also not be performed on such small areas of 200 x 200 μm$^2$ with the equipment, which we access to.

Therefore, we now prepared and irradiated one large area ZnO single-crystalline samples (1 cm$^2$) with the same Ga ion implantation conditions for a better measurement access. We performed spectral ellipsometry measurements for the as-implanted situation at three different angle of incidence (50°, 60° and 70°) to extract the complex dielectric function. The Ga doped samples based on a layer stack model, which was used in previous publication on the same material and is explained in the supplementary information part of the work.[20] In **Figure S6**(a,b) we calculated for the Ga irradiated ZnO ($3.6 \times 10^{15}$ cm$^{-2}$) with the WVASE32 software and the extracted optical constants n&k of the Drude-Lorentz fit the spectral ellipsometry data $\Psi$ and $\Delta$ at two different angles of incidence, which are in good agreement with the measurements. The comparison of the extracted optical constants in **Figure S6**(c) are based on the calculated $\Psi$ and $\Delta$ data from the DL-fit (solid lines) and the measured SE data (dashed lines) in **Figure S6**(a,b). The comparison gives a good agreement for respective spectral range.

**Figure S6**(d) shows the compared reflectance, which was measured by FTIR (green curves) and calculated with the extracted n & k from **Figure S6**(c).

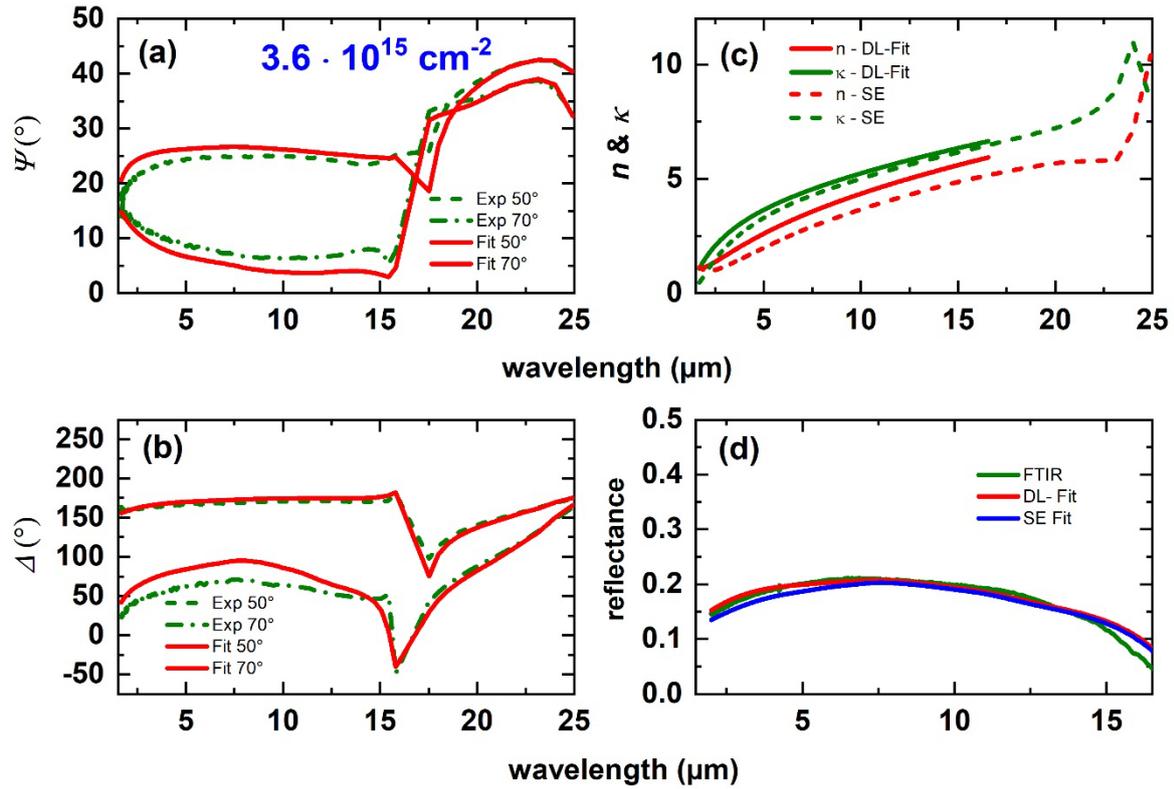

**Figure S6.** (a,b) Measured ellipsometry data (dashed lines) plus the calculated $\Psi$ and $\Delta$ (solid curves) from the Drude-Lorentz fit parameters for the unannealed Ga:ZnO samples, irradiated by a 30 keV ion accelerator with an ion fluence of $3.6 \times 10^{15}$ cm$^{-2}$ (corresponding to Ga peak concentrations of 3.1 at.%). (c) Comparison between the extracted complex refractive index of the unannealed Ga:ZnO layer from the Drude-Lorentz fit (solid lines) and the Spectral Ellipsometry fit (dashed lines) (d) Comparison between calculated reflectance with WVASE32 software and reflectance fitting parameters (red curve), the fitted reflectance by the measured SE data (blue curve) and the FTIR measurement (green curve) for the unannealed Ga:ZnO sample.

The same procedure in **Figure S6** was done for the highest ($6 \times 10^{15}$ cm$^{-2}$) Ga implanted ZnO samples in **Figure S7**.

Overall the investigated comparison between the optical constants from the DL-fit at reflectance measurements and model fit at the VASE measurements highlights a good agreement for the used ion fluence range for the unannealed and Ga doped ZnO samples. Thus we are expecting a similar good matching for the laser annealed samples, which allows us to extract the optical constants just by reflectance measurements.

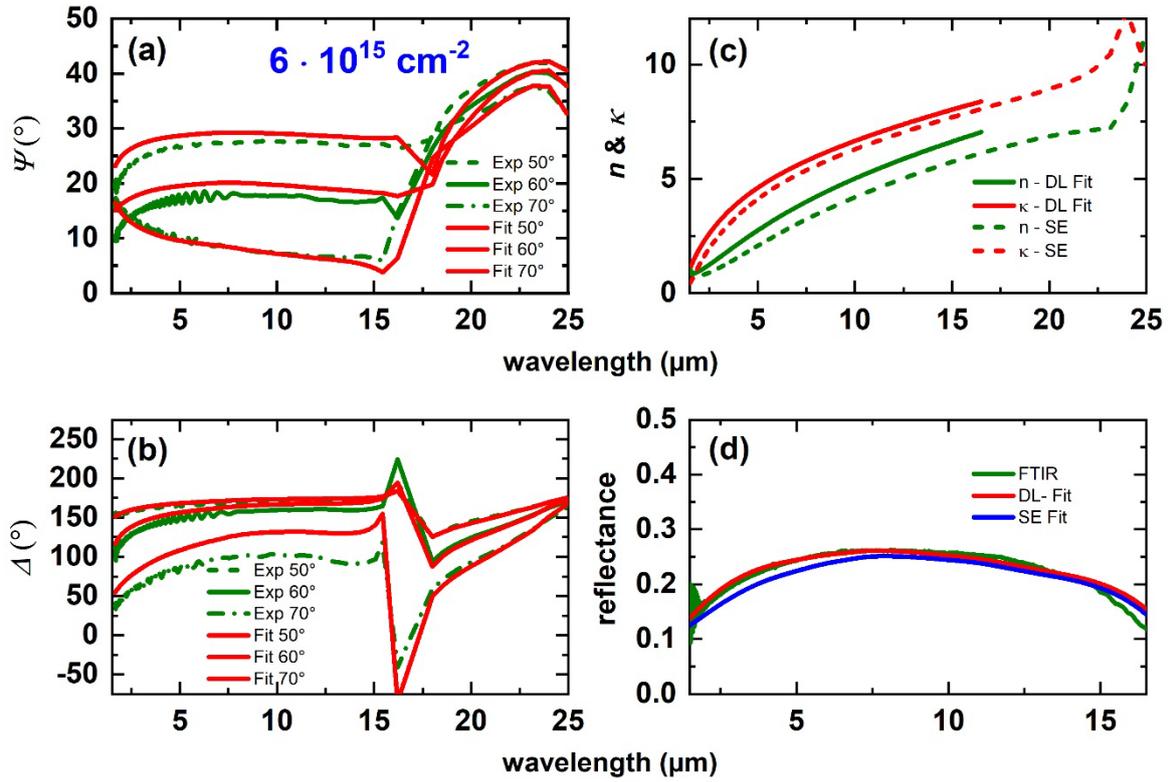

**Figure S7.** (a,b) Measured ellipsometry data (dashed lines) plus the calculated Ψ and Δ (solid curves) from the Drude-Lorentz fit parameters for the unannealed Ga:ZnO samples, irradiated by a 30 keV ion accelerator with an ion fluence of 6 × 10$^{15}$ cm$^{-2}$ (corresponding to Ga peak concentrations of 5.2 at.%). (c) Comparison between the extracted complex refractive index of the unannealed Ga:ZnO layer from the Drude-Lorentz fit (solid lines) and the Spectral Ellipsometry fit (dashed lines) (d) Comparison between calculated reflectance with WVASE32 software and reflectance fitting parameters (red curve), the fitted reflectance by the measured SE data (blue curve) and the FTIR measurement (green curve) for the unannealed Ga:ZnO sample.

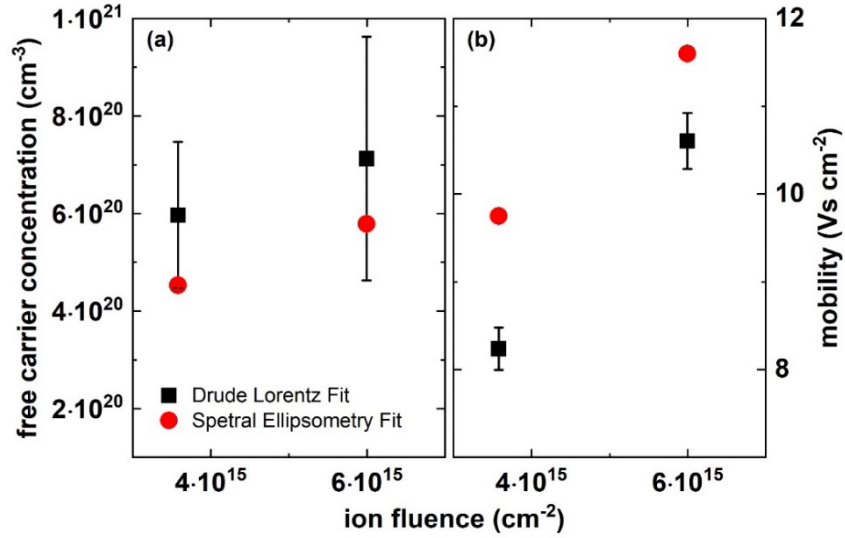

**Figure S8.** (a,b) Comparison between activated free carrier concentration and electron mobility from heavily Ga irradiated ZnO samples (ion fluence range $3.6 \times 10^{15} - 6 \times 10^{15}$ cm$^{-2}$), which were extracted from a Drude-Lorentz fit of reflectance measurements (black squares) and the spectral ellispometry measurements (red circles).

**Figure S8**(a,b) shows the comparison between the free carrier concentrations and mobilities based on the Drude-Lorentz fit (DL) of the reflectance (black rectangles) and the spectral ellipsometry (SE) measurements (red circles) on heavily Ga irradiated ZnO samples without annealing (ion fluence range $3.6 \times 10^{15} - 6 \times 10^{15}$ cm$^{-2}$). The free carrier concentrations from the DL in **Figure S7**(a) are close together to the SE fitted values and in the measurements error range. Vice versa is the situation for the mobilities of our Ga as implanted ZnO in **Figure S8**(b), where $\mu$ values for SE are little bit higher. The electron mobilities were calculated from the Drude damping term $\Gamma_D$ with: $\mu \sim 1/\Gamma_D$.

The good match between calculated $\Psi$ and $\Delta$ spectral ellipsometry data, for two or three different angles of incidence from only one reflectance measurement, and the SE measurements is expected. Thus, the Drude-Lorentz fitting routine of reflectance measurements serves as an easy and solid way to determine the optical constants of Ga doped ZnO.